\documentclass[doublespaced,a4paper,12pt]{article}
\usepackage[polish, english]{babel}
\usepackage[cp1250]{inputenc}
\usepackage{amsmath,amsmath,graphicx,epsf,pstricks,amsfonts,epsfig}
\title{\bf Vacuum driven accelerated expansion\footnote{Devoted to the memory of professor Ryszard R\c{a}czka on
the occasion of the 12-th anniversary of his death.}}

\author{Bogus{\l}aw Broda\footnote{bobroda@uni.lodz.pl}, Piotr
Bronowski\footnote{bronowski@gmail.com}, Marcin
Ostrowski\footnote{m.ostrowski@merlin.phys.uni.lodz.pl} and
Micha{\l} Szanecki\footnote{michalszanecki@wp.pl}\\
\small\textit{Department of Theoretical Physics,}
\small\textit{University of {\L}\'od\'z}\\
\small\textit{Pomorska 149/153, 90-236 {\L}\'od\'z, Poland}}
\date{}
\begin{document}
\maketitle
\begin{quote}
\noindent
\begin{flushleft}
\vspace{0.2cm} \textbf{PACS} 95.36.+x,
04.62.+v, 
98.80.Es\\
\end{flushleft}
\vspace{0.5cm} It has been shown that an improved estimation of
quantum vacuum energy can yield not only theoretically acceptable
but also experimentally realistic results. Our idea consists in a
straightforward extraction of gravitationally interacting part of
the full quantum vacuum energy by means of gauge transformations.
The implementation of the idea has been performed in the euclidean
version of the formalism of effective action, in the language of
Schwinger's proper time and the Seeley--DeWitt heat kernel
expansion, in the background of the
Friedmann--Robertson--Walker--Lema\^{\i}tre geometry.
\end{quote}

\section*{\hspace{0.4cm}1\;\;Introduction}
\noindent There are three famous problems in modern physics and
cosmology, which can, in principle, be treated as independent ones
or, just the opposite, (all or any two of them) as mutually
related:
\begin{flushleft}
 \begin{itemize}
 \item accelerated expansion of the Universe \cite{autor1,autor2} (proven by astrophysicists);
 \item cosmological constant or dark energy (very small, though
 non-vanishing) \cite{autor3,autor4};
 \item quantum vacuum energy density (theoretically --- very huge,
experimentally --- very small) \cite{autor5,autor6}.
 \end{itemize}
 \end{flushleft}

 The accelerated expansion of the Universe is by now rather a
 well-established by astronomical observations fact, in
 particular,
 Supernovae Ia data \cite{autor1}. That mysterious phenomenon still awaits an
 explanation. There are dozens of candidates for the solution of
 the problem (a number of approaches is reviewed in \cite{autor2}). One of the possible
 solutions and, in principle, the simplest one is the introduction of the cosmological constant $\Lambda$. Another solution (or the same, it depends on the point of view) is
 quantum vacuum energy \cite{autor5}. That solution is, in a sense, traditional because it seems to be theoretically the most natural and
 simple one, and it was proposed long ago. Its ``only drawback'' is the fact that, as it seems,
 it does not work well.

The cosmological constant problem \cite{autor3} troubles
 physicists from nearly the very beginning of the existence of
 general relativity. There are also dozens of candidates for the solution
 of this problem (they are even catalogued in \cite{autor7}).
 Unfortunately, explanation of the accelerated expansion by
 the vanishingly small value of the cosmological constant shifts
 only the problem rather than solves it.

 Traditional approach to
 the issue of the cosmological constant $\Lambda$ uses quantum
 vacuum energy as a source of the origin of that quantity. But still the mechanism,
 being very appealing, does not work, as it seems, properly. It
 appears that the traditionally calculated, Casimir-like value
 of quantum vacuum energy density is definitely too big than accepted, and two orders of orders too big than required!
 Entirely independently of the problem of the accelerated expansion and of
 the problem of the cosmological constant such a drastically huge
 value of the vacuum energy density is a serious problem in itself.
 That means that one should solve the quantum vacuum energy problem
 independently whether it could or should be later related to the
 accelerated expansion and the cosmological constant or not. There
 is quite a numerous collection of potential explanations of the
 above issues in literature. They are mainly given in the context of the dark energy \cite{autor2}. It is not our intention to list
 or review them but it seems to be useful for our
 further purposes to mention at least some.

One of early standard ideas was to lower the ultraviolet cutoff
scale $\Lambda_{\rm \textsc{uv}}$ using supersymmetry arguments.
It helps a bit, but only a little bit, if we want to be in
accordance with current experimental facts (roughly, it cuts the
order of discrepancy by two \cite{autor3}).

As another, rather a radical solution one should mention the idea
assuming that quantum vacuum energy does not, for one or another
reason, influence on gravity. For example, the authors of
\cite{autor8}, referring to the Casimir effect, rule out
the possibility that the observed cosmological constant arises
from the zero-point energy which is made finite by a suitable
cut-off. In \cite{autor6} it is claimed that the huge contribution
of the zero-point motion of the quantum fields to the vacuum
energy is exactly cancelled by the higher-energy degrees of
freedom of the quantum vacuum (automatic compensation of
zero-point energy). Or simply, ``zero-point energy does not
gravitate in vacuum''. And finally, the author of the paper
\cite{autor15} provocatively asks: ``...Why the vacuum does not
gravitate ...''?

It appears, and the aim of our paper is to show it, that, in
principle, it is possible to reasonably estimate the value of
quantum vacuum energy obtaining an experimentally acceptable
result. Moreover, the result is not only realistic in itself (the
quantum vacuum energy is not huge) but experimentally expected as
well. Our approach does not appeal to any more or less clever,
arbitrary or exotic assumption, and this is, in our opinion, its
main advantage. Just the opposite, our idea is supposed to adhere
to standard quantum field theory formalism as closely as possible.
In fact, our proposal only works provided standard philosophy of
quantum field theory is carefully taken into account.

\section*{\hspace{0.4cm}2\;\;Quantum vacuum energy}
The well-known, standard (but, regrettably, not properly working)
approach to estimation of the quantum vacuum energy density
$\varrho_{\rm vac}$ calculates the Casimir-like energy density for
the whole Universe. The result of such a calculation for a single
bosonic scalar mode (in mass units) is \cite{autor3}
\begin{align}
  \varrho_{\rm vac}=\frac{1}{2}\int\limits_{0}^{\Lambda_{\rm \textsc{uv}}}
  \frac{4\pi}{(2\pi\hbar)^3c}\;
  \sqrt{(mc)^2+k^2}\;k^2\mathrm{d}k,
  \label{eq:wrong density 1}
\end{align}
where $m$ is the mass of the mode. For a large ultraviolet (UV)
momentum cutoff $\Lambda_{\rm \textsc{uv}}$
\begin{align}
  \varrho_{\rm vac}\approx\frac{1}{(4\pi)^2}\frac{{\Lambda_{\rm \textsc{uv}}}^4}{\hbar^3
  c}.
  \label{eq:wrong density 2}
\end{align}
Setting $\Lambda_{\rm \textsc{uv}}=\Lambda_{\rm P}$, where
$\Lambda_{\rm P}$ is the Planck momentum,
\begin{align}
 \Lambda_{\rm P}=\sqrt{\frac{\hbar c^3}{G}}\approx 6.5\rm\,kg\,m/s,
 \label{eq:Planck momentum measure}
\end{align}
here $G$ is the newtonian gravitational constant, we obtain
\begin{align}
 \varrho^{\rm P}_{\rm vac}\approx\frac{c^5}{(4 \pi)^2 \hbar G^2}\approx
 3.4\times10^{94}\rm\,kg/m^3,
 \label{eq:wrong density 3}
\end{align}
an enormously huge value, whereas the experimentally estimated
value is of the order of, the so-called, critical density of the
Universe $\varrho_{\rm crit}={3 {H_{0}}^{2}}/{8\pi G}$
($\approx10^{-26}\rm\,kg/m^3$), where $H_{0}$ is the present day
Hubble expansion rate, i.e.\ more than 120 orders less! Lowering
$\Lambda_{\rm \textsc{uv}}$ to, say, the supersymmetry scale
$\Lambda_{\rm\textsc{susy}}\approx1\rm\,TeV/c$, only slightly
improves the situation, namely, $\varrho_{\rm
vac}^{\rm\textsc{susy}}\approx 1.5\times10^{30}\rm\,kg/m^3$, but
it does not change the general impression that the whole
calculation is principally erroneous. Therefore, as a desperate
response to this dramatic situation, the earlier mentioned idea
has emerged that gravitational field is insensitive to quantum
vacuum fluctuations, yielding $\varrho^{0}_{\rm vac}=0$.

The both described, extreme approaches, the ordinary, purely
Casimir-like calculation and the insensitiveness idea actually
yield entirely, from experimental point of view, incorrect
results. That fact should become obvious also from theoretical
point of view for the following reasons. First of all, we observe
that the ordinary, purely Casimir-like calculation of quantum
vacuum energy should not give any measurable contribution to
gravitational (or any other) field ``by construction''. Actually,
any classically analysed process (interaction) is an approximation
of a quantum one. Therefore, we are allowed to reason in the
language of Feynman diagrams. Let us observe that the ordinary,
purely Casimir-like calculus gives rise to contributions coming
from closed loops (see, Fig.~1) without any external lines. They
do not influence gravitational field because this possibility has
not been taken into account by virtue of the construction, i.e.\
there are no ``classical'' external lines establishing contact of
the internal ``matter'' loops with the outer gravitational field.

\begin{center}
\includegraphics[scale=0.7]{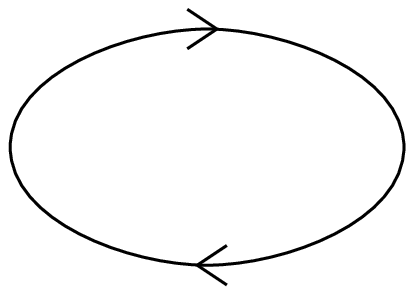}
\end{center}
\begin{center}
\emph{\textbf{Fig.1:} A single closed loop representing the
ordinary, purely Casimir-like contribution of a free matter field
to the effective action.}
\end{center}

But one can easily correct this trivial result performing improved
calculations. Namely, one should consider contributions coming
from closed ``matter'' loops with classical external gravitational
lines attached. Such an approach is not only in full accordance
with paradigms of standard quantum field theory, without any
additional assumptions, but also, moreover, it would bring us to
realistic results.

Summarizing, roughly speaking, we should calculate all quantum
vacuum fluctuations of a matter field in an external classical
gravitational field, retaining the most divergent parts (as it
will become clear in the next section), and next we should discard
the term without gravitational field.

In this section, we have sketched our idea of a proposed
estimation of the quantum vacuum energy. A concrete implementation
of the idea will be introduced in the following section. We
intentionally stress the difference between the idea and the
implementation because we believe that the proposed implementation
of the estimation is not final nor unique.
\section*{\hspace{0.4cm}3\;\;Implementation of the estimation}
Full quantum contribution coming from a single
(non-self-interacting) mode is included in the effective action of
the form \cite{autor9}
\begin{align}
 S_{\rm eff}=\pm\frac{\hbar}{2}\log\det\mathcal{D},
 \label{eq:EffectiveAction1}
\end{align}
where $\mathcal{D}$ is a non-negative, second-order differential
operator, in general, with classical external fields, and the
upper (plus) sign corresponds to bosonic statistics whereas the
lower (minus) one corresponds to fermionic statistics,
respectively. For simplicity, we work in the euclidean framework
throughout. Since \eqref{eq:EffectiveAction1} is UV divergent we
should regularize it, analogously to \eqref{eq:wrong density 1}.
The most convenient and systematic way to control infinities in
Eq.\,\eqref{eq:EffectiveAction1} is to use Schwinger's proper-time
method. In this approach \cite{autor9,autor13} we can formally
rewrite \eqref{eq:EffectiveAction1} as
\begin{align}
 S_{\rm
 eff}=\pm\frac{\hbar}{2}\mathrm{Tr}\log\mathcal{D}=\mp\frac{\hbar}{2}\int\limits_0^{\infty}\frac{\mathrm{d}s}{s}\;
 \mathrm{Tr}\;e^{-s\mathcal{D}}.
 \label{eq:EffectiveAction2}
\end{align}
Now, the UV regularized version of \eqref{eq:EffectiveAction2} is
\begin{align}
  S^{\varepsilon}_{\rm eff}=\mp\frac{\hbar}{2}\int\limits_{\varepsilon}^{\infty}\frac{\mathrm{d}s}{s}\;\mathrm{Tr}
  \;e^{-s\mathcal{D}}\equiv\mp\frac{\hbar}{2}\int\limits_{\varepsilon}^{\infty}\frac{\mathrm{d}s}{s}\;T_{\mathcal{D}}(s),
 \label{eq:EffectiveActionRegularized}
\end{align}
where $\varepsilon$ is an UV cutoff in the units: length to the
power two. Next, we apply the following Seeley--DeWitt heat kernel
expansion \cite{autor9,autor10}:
\begin{align}
T_{\mathcal{D}}(s)\equiv\int t(s;x)\;\sqrt{g}\;\mathrm{d}^4x\;,
 \label{eq:Seeley--DeWitt Expansion}
\end{align}
where, in four dimensions without boundary,
\begin{align}
t(s;x)=\frac{1}{(4\pi s)^2}\sum_{n=0}^{\infty}a_n(x)s^{n}.
 \label{eq:Seeley--DeWitt Expansion Cd}
\end{align}

A physical motivation to use the external lines approach is as
follows. A generally accepted and also our main assumption is the
statement that quantum vacuum fluctuations of ``matter'' (in fact,
all non-gravitational) fields somehow yield non-zero (effective)
cosmological constant (or dark energy). In other words, the vacuum
quantum fluctuations influence on gravitational field. In
technical terms, the influence, in the language of quantum field
theory, is established by lines of Feynman diagrams. Since we are
working in the domain of classical gravity, graviton lines are
supposed to be classical external lines, whereas (quantum)
``matter'' fields are confined to single loops.

The full expansion \eqref{eq:EffectiveActionRegularized}
corresponds to all one-loop Feynman diagrams for a matter field,
those depicted in Fig.~2 and also that in Fig.~1. But the
previously discussed, ordinary, purely Casimir-like vacuum diagram
in Fig.~1 should be discarded as a trivial one because, according
to our earlier discussion, by construction, it does not contain
any coupling to an external gravitational field. That trivial
vacuum diagram appears in the first coefficient of the
Seeley--DeWitt expansion \eqref{eq:Seeley--DeWitt Expansion Cd},
$a_0(x)$. It is obvious because only $a_0(x)$ survives the
vanishing external field limit. For any external gravitational
(and not only gravitational) non-vanishing or vanishing field, we
have $a_0(x)=1$.
\begin{center}
\includegraphics[scale=0.82]{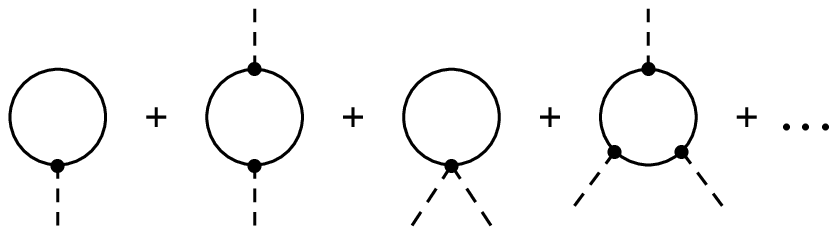}
\end{center}
\begin{center}
 \emph{\textbf{Fig.2:} Closed matter loops influencing gravitational field via attached classical external lines.}
\end{center}

Other $a_n(x)$'s (for $n>0$) contain various powers and
derivatives of curvature with dimensionality governed by $n$. In
particular, $a_1(x)=\frac{1}{6}R$ for an ordinary massless scalar
mode, where $R$ is the scalar curvature, and it finitely
renormalizes or induces \cite{autor11} (dependently on the point
of view) the classical Hilbert--Einstein action of gravity. For
that mode, the philosophy of the induced gravity yields, by virtue
of \eqref{eq:EffectiveActionRegularized}--\eqref{eq:Seeley--DeWitt
Expansion Cd},
\begin{equation}
\begin{split}
 S_{\rm ind} &=-\frac{\hbar}{2}\frac{1}{\varepsilon}\frac{1}{(4\pi)^2}\int\frac{1}{6}
 R\sqrt{g}\;\mathrm{d}^4x\\
 &=-\frac{\hbar}{12}\frac{1}{{L_{\rm P}}^2}\frac{1}{(4\pi)^2}\int
 R\sqrt{g}\;\mathrm{d}^4x\\
 &=-\frac{1}{12\pi} \frac{c^3}{16\pi G}\int
 R\sqrt{g}\;\mathrm{d}^4x,
\end{split}
\label{eq:Induced Action 1}
\end{equation}
where the planckian UV cutoff has been assumed, i.e.\
\begin{align}
\varepsilon={L_{\rm P}}^2=\frac{\hbar G}{c^3}. \label{eq:Cutoff of
Induced Action}
\end{align}
Hence, amazingly, the induced coupling constant for a single mode
is approximately $12\pi$ times less than the standard classical
value! The next term, $a_2(x)$, and also further terms, yields
genuine quantum corrections to the classical theory, and thus is
uninteresting for us.

As is commonly expected, the (effective) cosmological constant
$\Lambda$ or dark energy can be induced by the zeroth term, $a_0$,
and therefore we will concentrate on that term henceforth. The
zeroth term $a_0$ yields, according to
\eqref{eq:EffectiveActionRegularized}--\eqref{eq:Seeley--DeWitt
Expansion Cd} and \eqref{eq:Cutoff of Induced Action},
Casimir-like contribution of the form
\begin{equation}
\begin{split}
 S_{\rm Cas} &=\mp\frac{\hbar}{2}\frac{1}{2\varepsilon^2}\frac{1}{(4\pi)^2}\int \sqrt{g}\;\mathrm{d}^4x\\
 &=\mp\frac{\hbar}{4}\frac{1}{{L_{\rm P}}^4}\frac{1}{(4\pi)^2}\int \sqrt{g}\;\mathrm{d}^4x\\
 &=\mp\frac{1}{4} \frac{c^6}{(4\pi)^2\hbar G^2}\int
 \sqrt{g}\;\mathrm{d}^4x.
\end{split}
\label{eq:Casimir Action 1}
\end{equation}
In the flat space limit it corresponds (in mass density units) to
the value exactly 4 times less than that calculated earlier in
\eqref{eq:wrong density 3}. This unessential difference is coming
from different regularization procedures. Besides, we should
remember that hamiltonian and lagrangian are different objects.
Since, by assumption, we only perform estimations we may not care
of that difference, and concentrate only on orders of values. In a
sense, we have derived a covariant version of the result
\eqref{eq:wrong density 3}. According to our strategy we have to
extract from Eq.\ \eqref{eq:Casimir Action 1} only the part
corresponding to gravitational field.

For technical simplicity, but in accordance with experimental
realm, we can take the metric of the spatially flat
Friedmann--Robertson--Walker--Lema\^{\i}tre (FRWL) form with the
scale factor $a(t)$,
\begin{equation}
g_{\mu\nu}=\left(%
\begin{array}{cccc}
  1 & 0 & 0 & 0 \\
  0 & a^{2}(t) & 0 & 0 \\
  0 & 0 & a^{2}(t) & 0 \\
  0 & 0 & 0 & a^{2}(t) \\
\end{array}%
\right).
\label{eq:FRWL Metric Matrix Form}
\end{equation}
For calculational purposes, let us assume that the coordinate time
$t=0$ corresponds to the present moment, and normalize the
coordinates according to the equality
\begin{equation}
a(0)=1.
\label{eq:Scale factor limit}
\end{equation}
Now, power-series expanding around $t=0$, we have
\begin{equation}
\begin{split}
a(t) &=a(0)+\dot{a}(0)t+\ddot{a}(0)t^{2}+O(t^3)\\
&=1+H_0t-\frac{1}{2}q_0{H_0}^2t^2+O(t^3),
\end{split}
\label{eq:Metric Power Expansion}
\end{equation}
where the present day Hubble expansion rate,
$$H_0\equiv\frac{\dot{a}(0)}{a(0)}=\dot{a}(0),$$ and $q_0$ is the present day deceleration
parameter,
$$q_0\equiv-{H_0}^{-2}a^{-1}(0)\ddot{a}(0)=-{H_0}^{-2}\ddot{a}(0).$$ Hence
\begin{equation}
\sqrt{g}=\left[a^2(t)\right]^\frac{3}{2}=\left[1+2H_0t+(1-q_0){H_0}^2t^2+O(t^3)\right]^\frac{3}{2}.
\label{eq:Metric Measure}
\end{equation}

It is of vital importance for our further considerations to show
that the second term in \eqref{eq:Metric Measure}, linear in $t$,
can be discarded by virtue of gauge symmetry. A general,
model-independent discussion of this fact is given in
\cite{autor14}. Physically, such a potential possibility
corresponds to the obvious fact that not any perturbation of flat
metric represents a genuine gravitational field but only that
which is gauge nontrivial. The direct proof goes as follows.
Infinitesimal gauge transformations around flat metric are given
by
\begin{equation}
\delta g_{\mu\nu}=\partial_{\mu}\xi_{\nu}+\partial_{\nu}\xi_{\mu},
\label{eq:Gauge transformation 1}
\end{equation}
where $\xi_{\mu}=\left(\xi\equiv\xi_{0},\;\xi_i\right)$ are gauge
parameters. Explicitly, the first equation is
\begin{equation}
\delta g_{00}=2\dot{\xi}=0,
\label{eq:Gauge equation 1}
\end{equation}
because $g_{00}=1$ should be left undisturbed. A general solution
of Eq.\ \eqref{eq:Gauge equation 1} is then $\xi=\xi(\bold{x})$.
The second equation is of the form
\begin{equation}
\delta g_{0i}\equiv\delta g_{i0}=\dot{\xi_i}+\partial_{i}\xi=0,
\label{eq:Gauge equation 2}
\end{equation}
because also $g_{0i}=0$ should be left intact. Hence
$\dot{\xi_i}=-\partial_{i}\xi(\bold{x})$, and consequently
\begin{equation}
\xi_i=-t\;\partial_{i}\xi(\bold{x})+\eta_i(\bold{x}).
\label{eq:Gauge equation 3}
\end{equation}
For purely spatial indices we have
\begin{equation}
 \begin{split}
\delta g_{ij}=&\partial_{i}\xi_j+\partial_{j}\xi_i\\
=&-2t\;\partial_{ij}\xi(\bold{x})+\partial_i\eta_j(\bold{x})+
\partial_j\eta_i(\bold{x}).
\end{split}
\label{eq:Gauge equation 4}
\end{equation}
Now, we put
\begin{equation}
\delta g_{ij}=\left\{
\begin{array}{ccl}
  0, & \text{for} & i\neq j\\
 f(t), & \text{for} & i=j.
\end{array}
\right.
\label{eq:Gauge equation 5}
\end{equation}
From Eq.\ \eqref{eq:Gauge equation 4} it immediately follows that
the most general function $f(t)$ which can be gauged away is
linear in $t$. As a final, sufficiently general solution of our
equation we assume the particular form
\begin{equation}
\xi(\bold{x})=\frac{1}{2}\sum\limits_{i,j=1}^3\xi_{ij}x^ix^j,\;\;\eta(\bold{x})=0,
\label{eq:Gauge equation 6}
\end{equation}
where the constant matrix $\xi_{ij}$, in view of Eq.\
\eqref{eq:Gauge equation 5}, should be diagonal (even scalar),
i.e.\ $\xi_{ij}=\frac{1}{2}H_0\delta_{ij}$, after appropriate
normalization. Then, the solution of our problem is of the form
\begin{equation}
\xi_{\mu}=\left(\frac{1}{4}H_{0}{\bold{x}}^2,
-\frac{1}{2}H_{0}tx^{i}\right). \label{eq:KSI solution}
\end{equation}
Gauging away the term linear in $t$ in Eq.\ \eqref{eq:Metric
Measure}, using \eqref{eq:KSI solution}, we have for small $t$
\begin{equation}
\sqrt{g}=1+\frac{3}{2}(1-q_0){H_0}^2t^2+O(t^3). \label{eq:Metric
Measure Approx}
\end{equation}
Since the integrand in Eq.\ \eqref{eq:Casimir Action 1} is only
$t$-dependent we can painless divide it by the spatial volume
$\int\mathrm{d}^{3}x$. Instead, dividing by the time
coordinate, at least for small $t$, is nothing but an averaging
procedure with respect to $t$. As our analysis is perturbative in
the time $t$, the longer the time the smaller the reliability of
our analysis. The shortest possible time, in the realm of quantum
field theory, is $t=T_{\rm P}$ (the Planck time). Therefore, time
averaging ${\langle\,\cdot\,\rangle}_t$ around present moment
$(t=0)$ is given by the formula
\begin{equation}
{\langle\,\cdot\,\rangle}_t\equiv\lim_{T\rightarrow T_{\rm
P}}\frac{1}{T}\int \limits_0^T \mathrm{d}t\;(\;\cdot\;).
\label{eq:Time Averaging}
\end{equation}
Thus, an estimated density is according to Eq.\ \eqref{eq:Casimir
Action 1} of the order
\begin{equation}
\begin{split}
\varrho=\mp\frac{1}{4}\frac{c^5}{(4\pi)^2\hbar
G^2}\lim_{T\rightarrow T_{\rm P}}\frac{1}{T}\int \limits_0^T
\mathrm{d}t\;(\sqrt{g}-1)\\
\approx \mp\frac{1}{4}\frac{c^5}{(4\pi)^2\hbar
G^2}\frac{1}{2}(1-q_0){H_0}^2{T_{\rm P}}^2,
\end{split} \label{eq:Good
Density 1}
\end{equation}
where the substraction in the parentheses in the first term
corresponds to discarding gravitationally non-interacting part of
the effective action. Equivalently, in Eq.\
\eqref{eq:EffectiveAction1} one should consider
$\det\mathcal{D}/\det\mathcal{D}_0$ instead of $\det\mathcal{D}$,
where $\mathcal{D}_0$ is a flat space version of $\mathcal{D}$. Or,
in another language, the subtraction corresponds to normalization
of the functional measure. Therefore, the subtraction is by no
means, as it could seem, an {\em ad hoc} procedure. Since ${T_{\rm
P}}^2=\hbar G/c^5$, and ${H_0}^2=\frac{8}{3}\pi G \varrho_{\rm
crit}$, we finally obtain
\begin{equation}
\varrho\approx\mp \frac{1}{48\pi}(1-q_0)\varrho_{\rm crit}.
\label{eq:Good Density 2}
\end{equation}
Amazingly, Eq.\ \eqref{eq:Good Density 2} predicts a highly
reasonable result. Inserting $q_0\approx-0.7$, which is
phenomenologically a realistic assumption \cite{autor12}, yields
the following numerical result:
\begin{equation}
\varrho\approx\mp 0.01\;\varrho_{\rm crit}. \label{eq:Good Density
3}
\end{equation}
The experimental value is roughly $\varrho_{\rm exp}\approx
0.76\;\varrho_{\rm crit}$, therefore we finally have
\begin{align}
\varrho\approx\mp0.01\varrho_{\rm exp}
\label{eq:Final Density}
\end{align}
per a single mode, which is a very good estimation in our opinion.
Taking into account the remarks directly following Eq.\
\eqref{eq:Casimir Action 1}, $\varrho$ could be just as well
$0.04\;\varrho_{\rm exp}$. We should bear in mind that our
analysis is an estimation, and we are only interested in the
orders of values.

\section*{\hspace{0.4cm}4\;\;Conclusions}
Using coordinate gauge freedom we have managed to extract from the
full quantum vacuum term induced in an external classical
gravitational background by a fluctuating mode of a matter field
the fraction corresponding to interaction with gravitational
field. An explicit calculus has been performed in the framework of
the spatially flat FRWL geometry. The value of the contribution
coming from a single mode, which appears to be of the order of one
hundredth of the experimentally expected value, seems to be very
promising. Thus, the old, primary expectation that quantum vacuum
fluctuations could be the source of the cosmological constant
$\Lambda$ does not seem to be unjustified.

Although, our idea, in its definite form (as well as the
derivation), is novel, actually it emerges in many papers as more
or less explicitly expressed thoughts. For example, the author of
\cite{autor15} claims that zero-point energy gravitates in some
environments and not in vacuum. The author of \cite{autor16} is
even closer to our point of view claiming that we should consider
the fluctuations in the vacuum energy density. Interestingly,
Fig.~1(a) of \cite{autor17} suits our point of view excellently.

Finally, we would like to comment on the celebrated Casimir effect
and its possible relation to our approach. In fact, it happens
that the Casimir effect is often referred to by many authors
(including us) in the context of quantum vacuum. In particular, it
is usually claimed that reality of quantum vacuum energy is proved
by experimental confirmation of the Casimir effect itself. 
Rather surprisingly, Casimir forces are correctly described by a crude
count in the spirit of the derivation of Eq.~\eqref{eq:wrong
density 1}, without any external lines present. Therefore, we have
an apparent paradox. On the one hand the ordinary Casimir-like
calculation yields reasonable results in the case of the very
Casimir effect. On the other hand it yields useless results in
the context of gravity. The resolution is simple. The ordinary
Casimir-like calculation is a simplified approach that
(perhaps accidentally) somehow works. It appears that a deeper approach to the Casimir effect
\cite{autor18} is in spirit similar to our considerations. In particular, it
uses external lines and a subtraction of the ordinary Casimir-like
vacuum diagram (compare Fig.~3 in \cite{autor18} and our Fig.~2).

We would like to stress that we do not claim that we have found a
final solution of the problem of the accelerated expansion. We are
aware that there are a lot of more or less sensible competing
proposals in that area. All of them, including ours, have some
drawbacks. Our approach should be just interpreted in this
context, as a voice in the discussion, indicating a direction for
a possible further study.

\section*{\hspace{0.4cm}Acknowledgments}
This work was supported in part by the Polish
Ministry of Science and Higher Education Grant
PBZ/MIN/008/P03/2003 and by the University of  {\L}\'od\'z grant.


\begin{thebibliography}{5}
\bibitem[1]{autor1}A.\ G.\ Riess et al.,
Observational evidence from supernovae for an accelerating
universe and a cosmological constant, Astronom.\ J.\ \textbf{116},
1009--1038 (1998).

S.\ Perlmutter et al., Measurements of $\Omega$ and $\Lambda$ from
42 high-redshift Supernovae, Astrophys.\ J.\ \textbf{517},
565--586 (1999).

\bibitem[2]{autor2}
E.\ J.\ Copeland, M.\ Sami, and S.\ Tsujikawa, Dynamics of dark
energy, Int.\ J.\ Mod.\ Phys.\ D \textbf{15}, 1753--1936 (2006);
hep-th/0603057.

\bibitem[3]{autor3}
S.\ Weinberg, The cosmological constant problem, Rev.\ Mod.\
Phys.\ \textbf{61}, 1--23 (1989).

\bibitem[4]{autor4}
S.\ M.\ Carroll, The cosmological constant, Liv.\ Rev.\ Rel.\
\textbf{4} (2001); astro-ph/0004075.

T.\ Padmanabhan, Cosmological constant---the weight of the vacuum,
Phys.\ Rep.\ \textbf{380}, 235--320 (2003); hep-th/0212290.

T. Padmanabhan, Dark energy: mystery of the millennium, AIP Conf.\
Proc.\ \textbf{861}, 179--196 (2006), Albert Einstein
Century International Conference; astro-ph/0603114.

\bibitem[5]{autor5}
Y.\ B.\ Zel`dovich, Cosmological constant and elementary
particles, JETP Lett.\ \textbf{6}, 316--317 (1967), translated
from Zh.\ Eksp.\ Teor.\ Fiz., Pis`ma Redaktsiyu, \textbf{6}
883--884 (1967).

\bibitem[6]{autor6}
G.\ E.\ Volovik, Vacuum energy: myths and reality, Int.\ J.\ Mod.\
Phys.\ D \textbf{15}, 1987--2010 (2006); gr-qc/0604062.

G.\ E.\ Volovik, Cosmological constant and vacuum energy, Ann.\
Phys.\ (Leipzig) \textbf{14}, 165--176 (2005); gr-qc/0405012.

\bibitem[7]{autor7}
S.\ Nobbenhuis, Categorizing different approaches to the
cosmological constant problem, Found.\ Phys.\ \textbf{36},
613--680 (2006); gr-qc/0411093.

\bibitem[8]{autor8}
G.\ Mahajan, S.\ Sarkar, and T.\ Padmanabhan, Casimir effect
confronts cosmological constant, Phys.\ Lett.\ B \textbf{641},
6--10 (2006); astro-ph/0604265.

\bibitem[9]{autor15}
J.\ Polchinski, The cosmological constant and the string
landscape, hep-th/0603249.

\bibitem[10]{autor9}
B.\ S.\ DeWitt, Quantum field theory in curved spacetime, Phys.\
Rep.\ \textbf{19}, 295--357 (1975).

B.\ DeWitt, The Global Approach to Quantum Field Theory,
(Clarendon Press, 2003).

\bibitem[11]{autor13}
K.\ Huang, Quarks, Leptons and Gauge Fields, (World Scientific,
1992), Chapt.\ 10.4.

P.\ Ramond, Field Theory: A Modern Primer, (Addison-Wesley, 2001),
Chapt.\ 3.

See also:

T.\ P.\ Cheng and L.\ F.\ Li, Gauge Theory of Elementary Particle
Physics, (Oxford University Press, 2000), Chapt.\ 6.4.

S.\ Weinberg, The Quantum Theory of Fields Volume 2: Modern
Applications, (Cambridge University Press, 1996), Chapt.\ 16.2.

\bibitem[12]{autor10}
R.\ D.\ Ball, Chiral gauge theory, Phys.\ Rep.\ \textbf{182},
1--186 (1989).

\bibitem[13]{autor11}
M.\ Visser, Sakharov's induced gravity: a modern perspective,
Mod.\ Phys.\ Lett.\ A \textbf{17}, 977--992 (2002); gr-qc/0204062.

\bibitem[14]{autor14}
I.\ L.\ Shapiro, Effective action of vacuum: semiclassical
approach, 0801.0216 [gr-qc], Chapt.\ 3.1.

\bibitem[15]{autor12}
J.\ M.\ Virey, P.\ Taxil, A.\ Tilquin, A.\ Ealet, C.\ Tao, and D.\
Fouchez, On the determination of the deceleration parameter from
Supernovae data, Phys.\ Rev.\ D \textbf{72}, 061302 (2005);
astro-ph/0502163.

\bibitem[16]{autor16}
T.\ Padmanabhan, Dark energy: the cosmological challenge of the
millennium, Curr.\ Sci.\ \textbf{88}, 1057--1068 (2005);
astro-ph/0411044.

\bibitem[17]{autor17}
R.\ Bousso, TASI lectures on the cosmological constant, 0708.4231
[hep-th].

\bibitem[18]{autor18}
R.\ L.\ Jaffe, The Casimir effect and the quantum vacuum, Phys.\
Rev.\ D \textbf{72}, 021301 (2005); hep-th/0503158.

\end{thebibliography}
\end{document}